# USE OF BESTEST PROCEDURE TO IMPROVE A BUILDING THERMAL SIMULATION PROGRAM

T. SOUBDHAN*, T. A. MARA*,* H. BOYER** and A. YOUNES

*Université Antilles-Guyane, Département de Physique, BP 159 Pointe-à-Pitre cedex Guadeloupe - France

**Laboratoire de Génie Industriel, Equipe Génie Civil 15 avenue René Cassin, BP 7151 97715 Saint-Denis Messag Cedex 9 La Réunion – France. email : mara@univ-reunion.fr

## ABSTRACT

Validation of building energy simulation programs is of major interest to both users and modellers. To achieve such a task, it is essential to apply a methodology based on a priori test and empirical validation. A priori test consists in verifying that models embedded in a program and their implementation are correct. this should be achieved before carrying out experiments. The aim of this report is to present results from the application of the BESTEST procedure to our code. We will emphasise the way it allows to find bugs in our program and also how it permits to qualify models of heat transfer by conduction.

## KEYWORDS

## DESCRIPTION

### BESTEST procedure

The International Energy Agency (IEA) sponsors a number of programs to improve the use and associated technologies of energy. The National Renewable Energy Laboratory (NREL) developed BESTEST ( Judkoff & Neymark, 1995) which is a method based on comparative testing of building simulation programs, on the IEA's behalf. The procedure consists of a series of test cases buildings that are designed to isolate individual aspects of building energy and test the extremes of a program. As the modelling approach is very different between codes, the test cases are specified so that input equivalency can be defined thus allowing the different cases to be modelled by most of codes. The basis for comparison is a range of results from a number of programs considered to be a state-of-art in United States and Europe.

### CODYRUN

In Reunion Island, we developed, for both design and research purposes, a multimodel and multizone building energy simulation software called CODYRUN (Boyer et al., 1998). Indeed, one of the most interesting aspects of this code, is to propose different models of heat transfer. Models of heat transfer by conduction through walls are based on thermal and electrical analogy. It is possible to choose between the simplified model (only two nodes by wall) or a more discretized one ( the number of nodes depends on the number of layers in the wall).

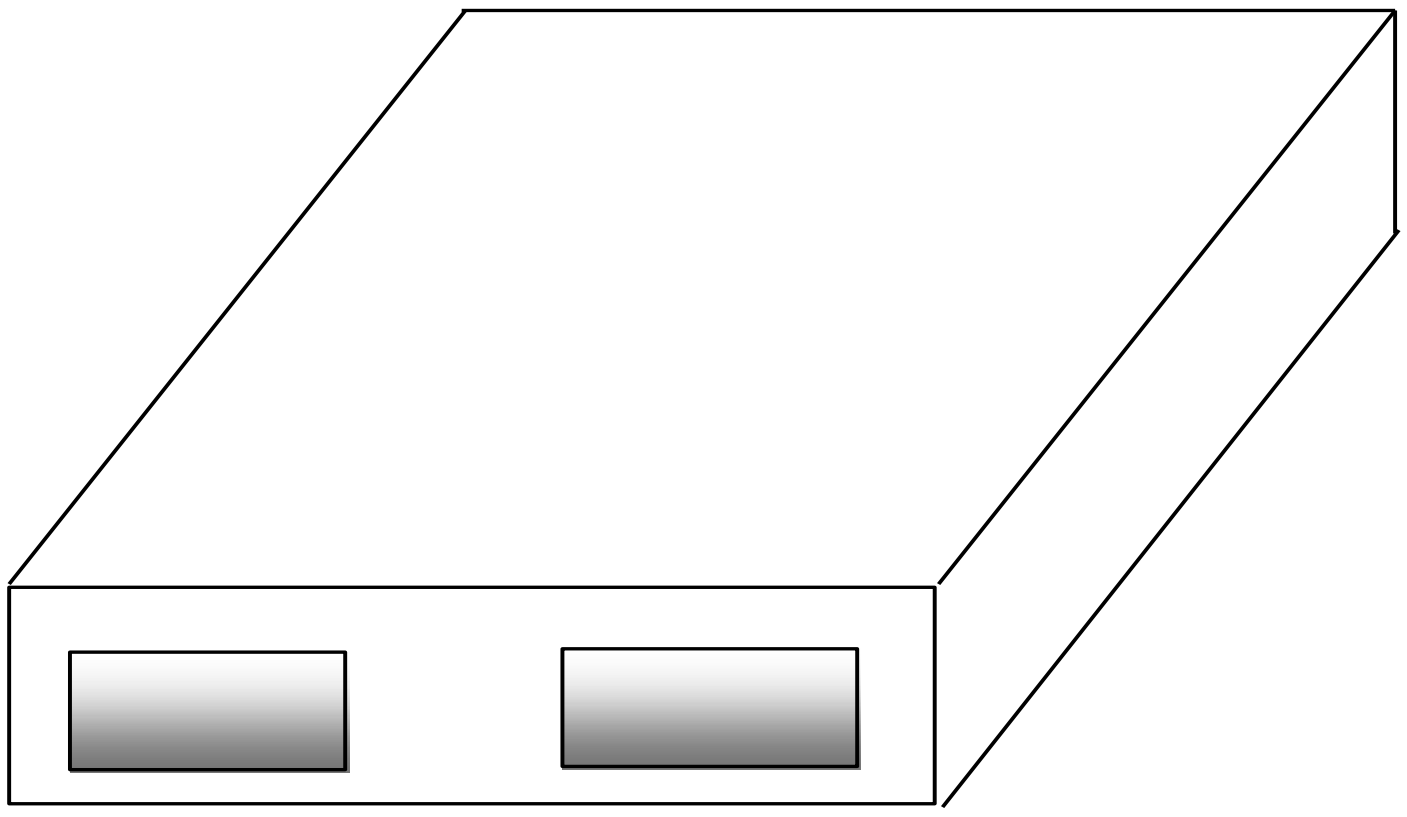

Fig. 1. Basic BESTEST building (case 600 & 900).

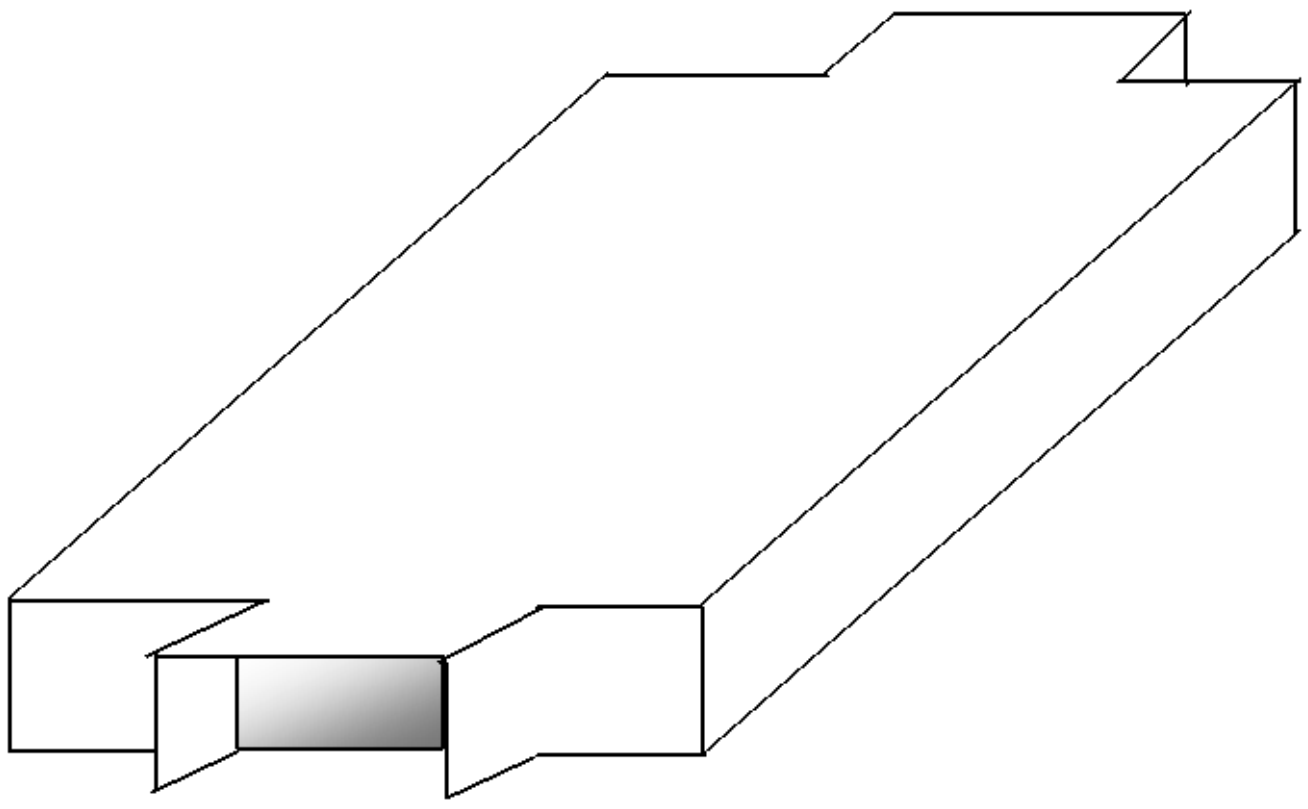

Fig. 2. Case 630 & 930, test East & West overhangs.

Implied in energy building simulation programs since 1994, we applied the BESTEST method to check our code. Indeed, even though CODYRUN's predictions were compared successfully to another code (Boyer, 1993) and to measurements (Garde, 1997) it seems to be more advisable to verify deeply its numerical implementation. BESTEST is an efficient method to trap bugs and faulty algorithms. Indeed it has already allowed to improve many codes.

## RESULTS

low-mass buildings

The first comparison (case 600) concerns a lightweight building with large windows on the south wall as shown in Fig. 1. It contains a perfect thermostat so that when the air temperature drops below 20°C, the thermostat heating is set on and when the air temperature exceeds 27°C, the thermostat cooling is set on. This case has been conceived with specific parametric values in the aim of testing mechanism and algorithms of solar transmission through the window. In regard to calculation time, the simplified model of heat transfer conduction is used.

Table 1. Comparison of predicted annual heating and cooling for case 600

| Programs (Countries) | ESP (UK) | BLAST (US/IT) | DOE2 (USA) | SRES/SUN (USA) | SERIRES (UK) | S3PAS (SPAIN) | TRNSYS (BEL/UK) | TASE (FINLAND) | CODYRUN (LA REUNION) |
|---|---|---|---|---|---|---|---|---|---|
| Annual Heating (MWh) | 4.296 | 4.773 | 5.709 | 5.226 | 5.596 | 4.882 | 4.872 | 5.362 | 3.850 |
| Annual Cooling (MWh) | 6.137 | 6.433 | 7.079 | 7.278 | 7.964 | 6.492 | 6.492 | 6.778 | 5.420 |

Analysis of the annual heating and cooling loads shows that CODYRUN disagrees with the other codes (see table 1) even though predicted incident and transmitted solar radiation seem identical (fig. 3). After having performed the diagnostic series, it appear that CODYRUN shows large disagreement with cases having extreme values for interior solar absorptance. This cases are case 270 and case 280 which allows to test the cavity albedo. The building is almost the same as the previous one except that, for case 270 the interior solar absorptance is 0.9 whereas for case 280 this parameter is equal to 0.1. We compare then the difference of annual heating and cooling consumption between those two models ( annual heating for case 280 minus annual heating for case 270, see fig. 4). The results observed here pinpoint a problem issued from the algorithm of the interior solar distribution. Fig. 4 shows clearly that the model of interior solar distribution is not accurate as CODYRUN overestimates the influence of interior solar absorptance on electrical consumption.

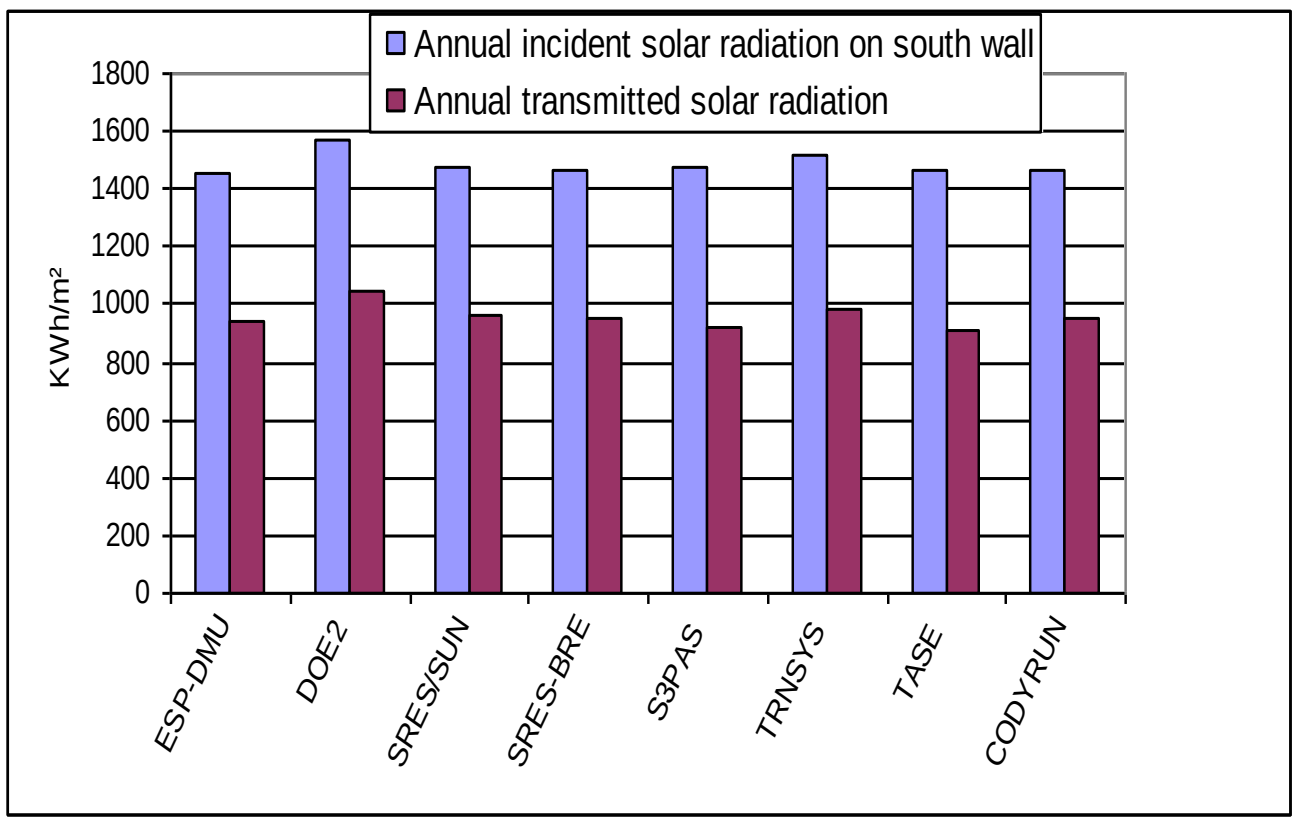

Fig. 3. Comparison of incident and transmitted solar radiation.

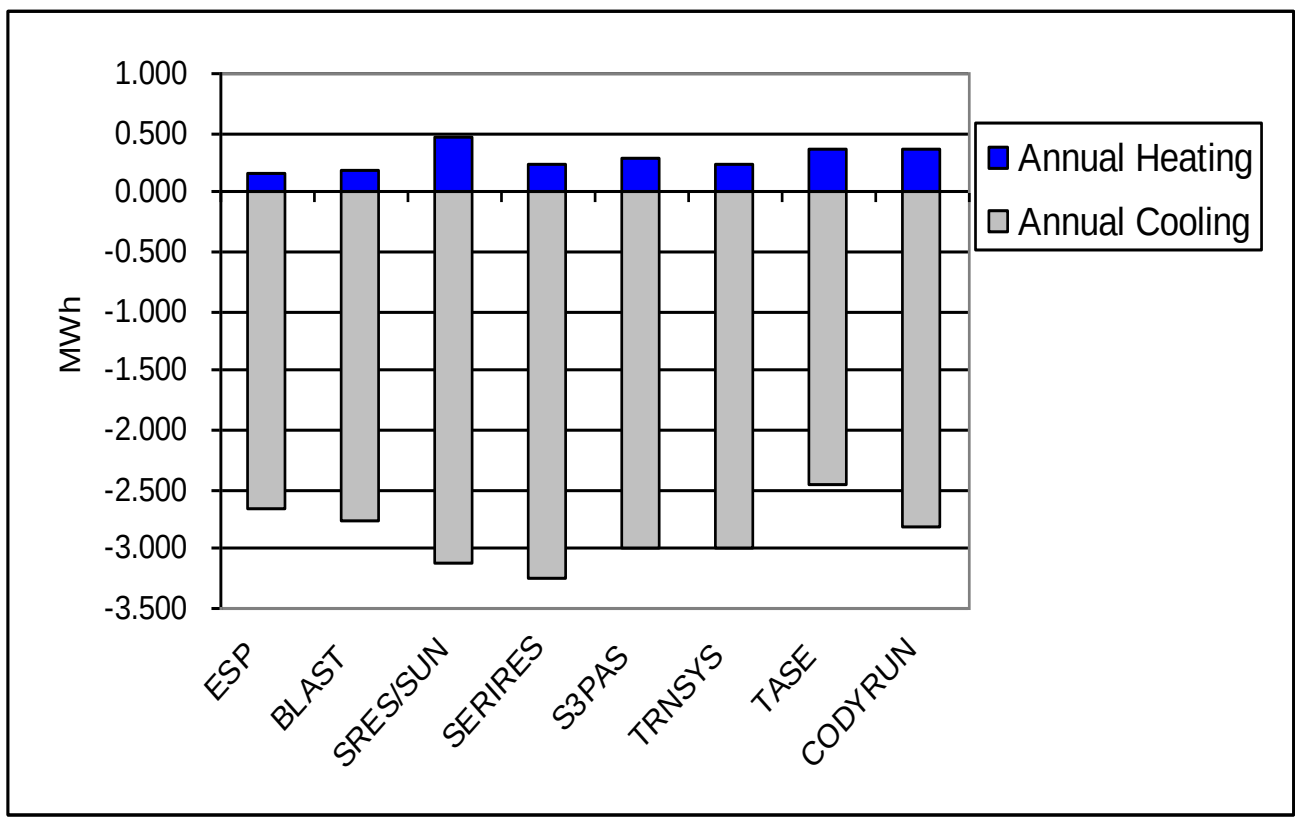

Fig. 5. Influence of interior solar absorptance on electrical consumption after improvement.

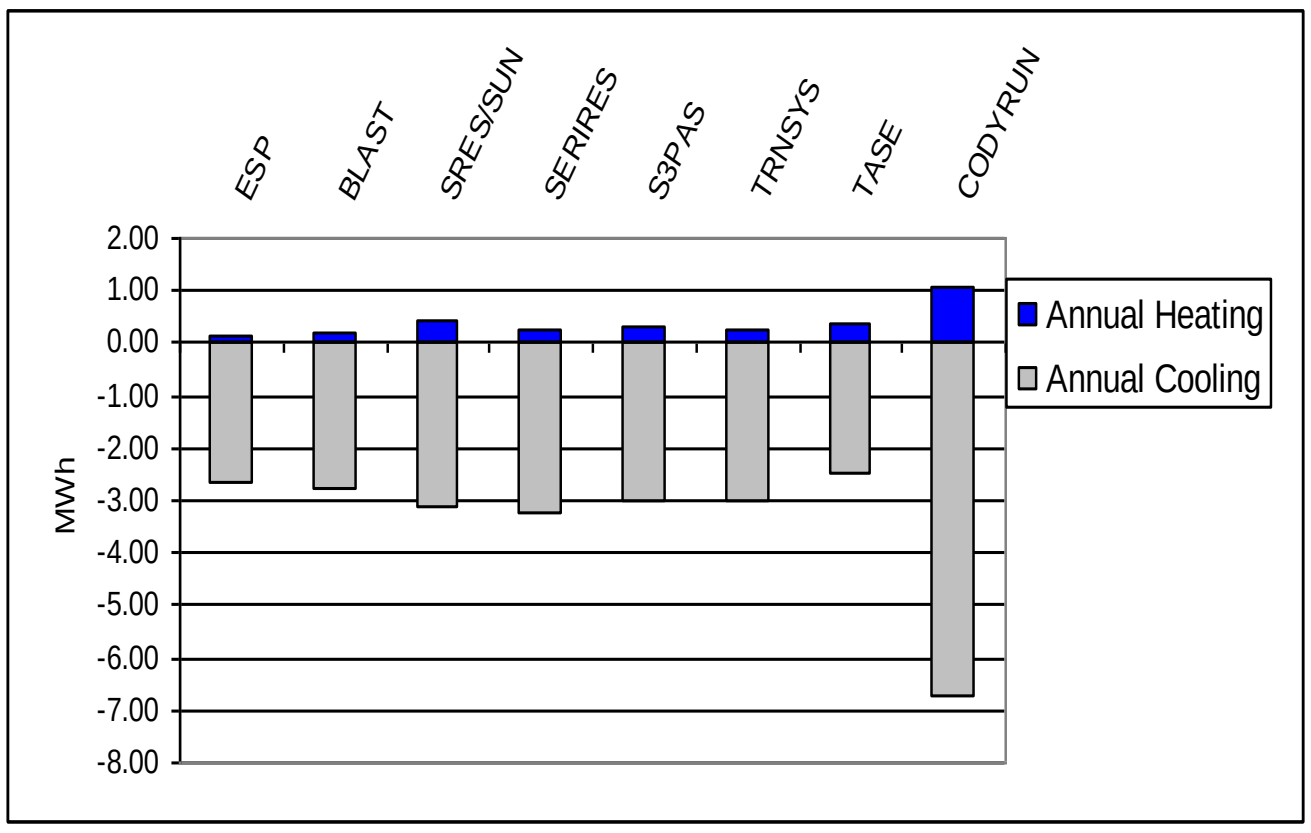

Fig. 4. Influence of interior solar absorptance on electrical consumption.

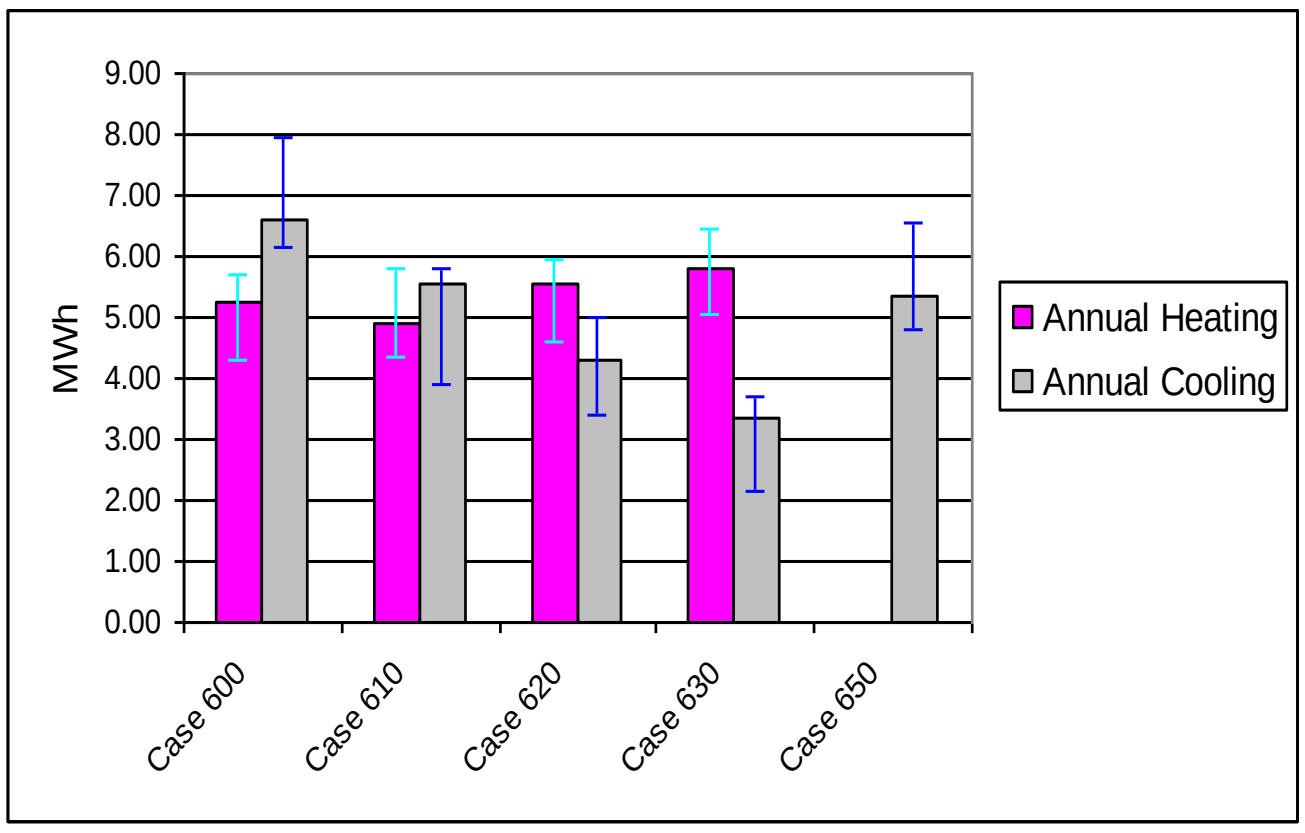

Fig. 6. Results of all lightweight building tests with their associated range.

<u>*Note*</u>
*Ranges associated to each case correspond to the minimal and maximal annual heating and cooling loads of the other programs.*
*Case 610 is the same as case 600 but adds a 1.0 m overhang on the south facing wall.*
*Case 620 is similar to case 600 except the windows are on the East and West facing walls.*
*Case 630 adds a wingwall and overhang to the geometry of case 620 around the windows.*
*Case 650 has the same geometry as case 600 with night venting used as well as cooling during the day. No heating is used.*

Consequently, we decided to check the implementation of the model of interior solar radiation fractions in the program. We found an error in the view factors calculation that lead to underestimate the fraction of solar radiation absorbed through the interior walls. The code was modified and tests 270 and 280 were performed once again. The influence of interior solar absorptance on electrical consumption, is then comparable to those predicted by the other codes (fig. 5). In fact, once this bug corrected, CODYRUN passes all the lightweight building tests successfully. Fig. 6 shows that the predicted results fall within the allowed range for each test.

<u>High-mass buildings</u>

The considered buildings are identical to the previous series of tests except that the walls are constituted of heavyweight materials. The first case (case 900) is similar to case 600 in the low-mass building tests and still as regard to calculation time, the simplified model of heat transfer conduction is used first. The model overestimates the electrical consumption (Table 2).The cause of discrepancies seems to be the heat transfer by conduction model's. Actually, the only difference between case 900 and case 600 is the walls constitution. To verify this hypothesis we use another model that discretizes a wall into seven nodes (see fig. 7).The predicted results with the new model are in good agreement with the other codes and CODYRUN passes all the heavyweight building tests successfully (fig. 8).

Table 2. Comparison of predicted annual heating and cooling for case 900 with the simplified model

| Programs | ESP | BLAST | DOE2 | SRES/SUN | SERIRES | S3PAS | TRNSYS | TASE | CODYRUN |
|---|---|---|---|---|---|---|---|---|---|
| Annual Heating (MWh) | 1.170 | 1.610 | 1.872 | 1.897 | 1.988 | 1.730 | 1.655 | 2.041 | 4.853 |
| Annual Cooling (MWh) | 2.132 | 2.600 | 2.455 | 3.165 | 3.415 | 2.572 | 2.485 | 2.599 | 6.184 |

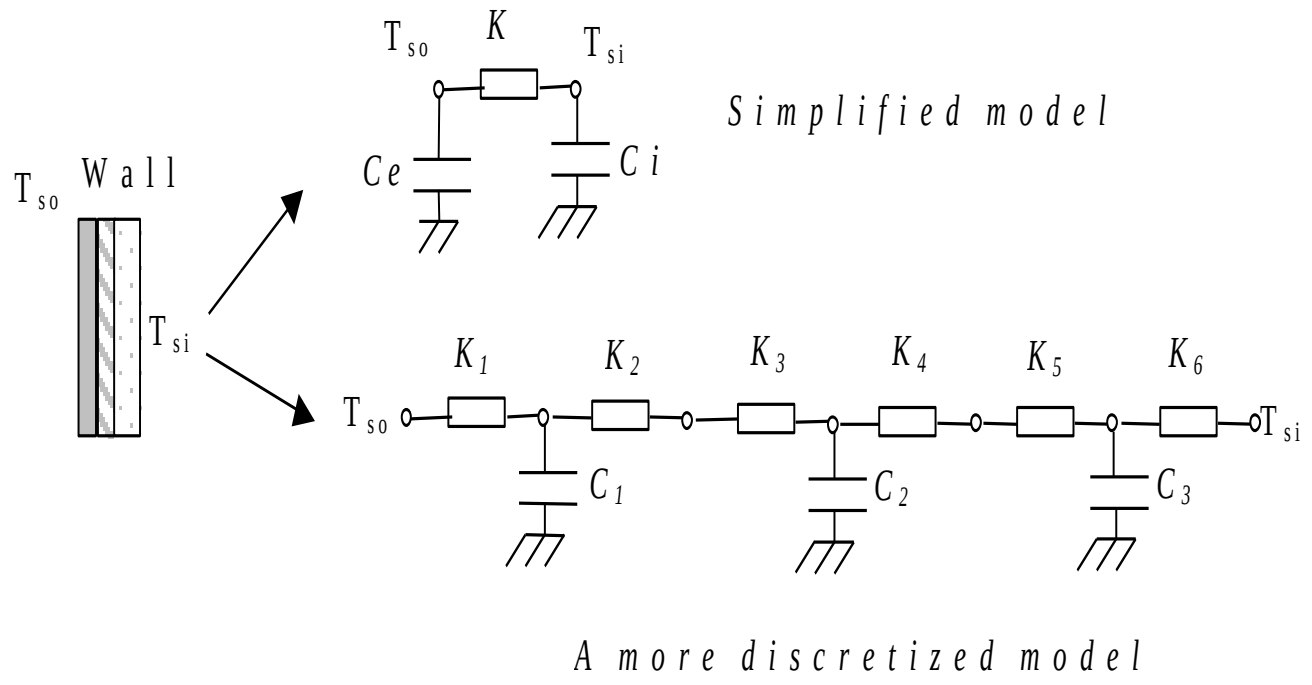

Fig. 7. Models of heat transfer by conduction

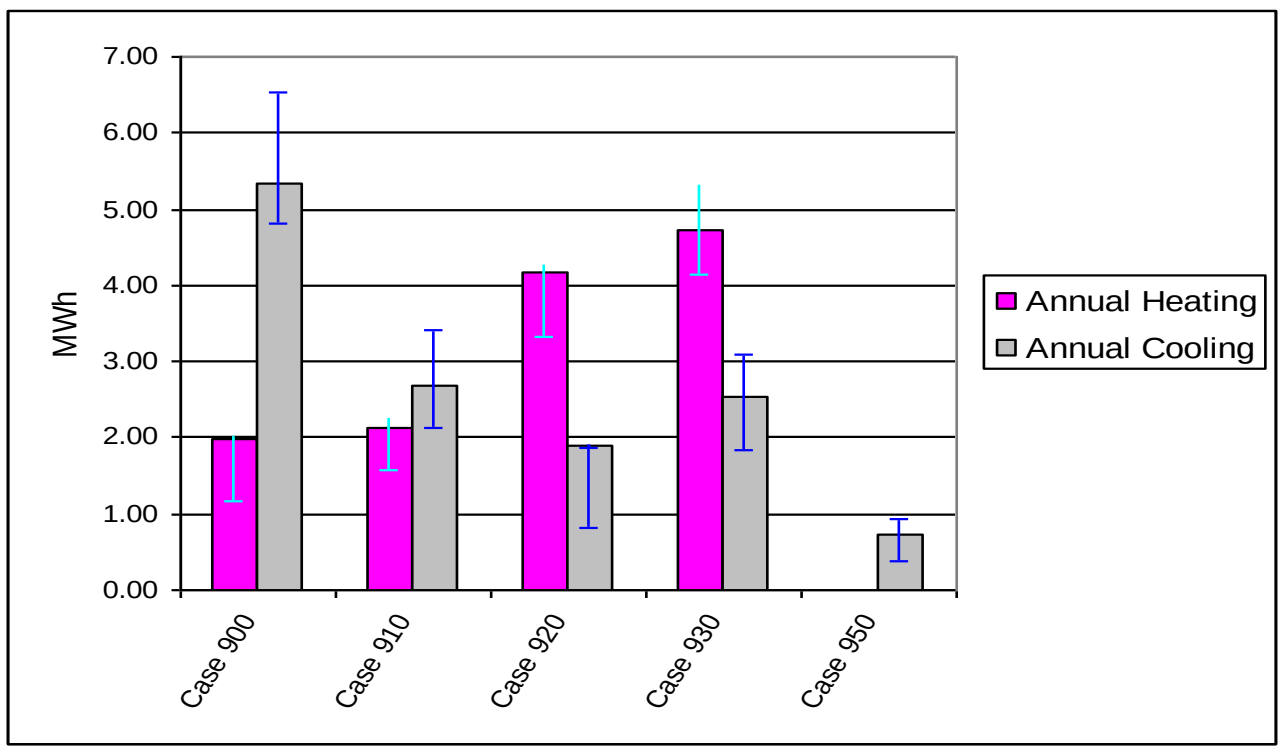

Fig. 8. Results of all heavyweight building tests with their associated range

This last information is very important as it means that the simplified model proposed by CODYRUN is unsuitable to simulate heat transfer by conduction through heavyweight walls. This can be explained by the fact that in this model, the thermal inertia is distributed on the wall's surfaces whereas in the seven node's model the thermal inertia is distributed inside the wall, which is more physically acceptable. In fact, this result even surprised developers and modellers who have never noticed this.

## CONCLUSION

Applying a comparative testing method to our tool revealed a certain number of problems that could not have been detected otherwise. It's a suitable method to check the numerical implementation of models in the program, but also for the verification of the validity range of different conduction models.
The BESTEST method is useful in validation of building energy simulation programs. It's an economic way of testing, in a few days, programs that have taken many years to develop. Our experience shows that the procedure not only allows to trap bugs in a program but also can help modellers to better understand models embedded in the software. Its not the perfect solution to the validation problem but is a big step made in this field.

## ACKNOWLEDGEMENT

The authors wish to thank Dr J. Neymark for his assistance and advice.